\documentclass[preprintnumbers,amsmath,amssymb,floatfix,10pt,prd,onecolumn,layout,superscriptaddress,nofootinbib]{revtex4}
\usepackage{latexsym}
\usepackage{epsfig}
\usepackage{epstopdf}
\usepackage{graphicx}
\usepackage{amssymb}
\usepackage{amsmath}
\usepackage{dcolumn}
\usepackage{bm}
\usepackage{color}
\usepackage{comment}
\usepackage{float}
\usepackage{array, makecell}
\usepackage[shortlabels]{enumitem}
\usepackage{subfigure}
\begin{document}
\title{\bf Holographic Einstein Rings of an AdS Black Hole in Massive Gravity}
\author{Xin-Yun Hu}
\altaffiliation{ hu\_xinyun@126.com}\affiliation{College of Economic
and Management, Chongqing Jiaotong University, Chongqing $400074$,
China}
\author{M. Israr Aslam}
\altaffiliation{mrisraraslam@gmail.com}
\author{Rabia Saleem}
\altaffiliation{rabiasaleem@cuilahore.edu.pk}\affiliation{Department
of Mathematics, COMSATS  University Islamabad, Lahore-Campus, Lahore
$54000$ Pakistan.}\author{Xiao-Xiong Zeng}
\altaffiliation{xxzengphysics@163.com}\affiliation{State Key
Laboratory of Mountain Bridge and Tunnel Engineering, Chongqing
Jiaotong University, Chongqing $400074$,
China}\affiliation{Department of Mechanics, Chongqing Jiaotong
University, Chongqing $400074$, China}
\begin{abstract}
In the context of holography, the Einstein ring of an AdS black hole (BH) in massive gravity (MG) is depicted. An oscillating Gaussian source on one side of the AdS boundary propagates in bulk, and we impose a response function to explain it. Using a wave optics imaging system, we obtain the optical appearance of the Einstein ring. Our research reveals that the ring can change into a luminosity-deformed ring or light spots depending on the variation of parameters and observational positions. When observers are positioned at the north pole, the holographic profiles always appear as a ring with concentric stripe surroundings, and a bright ring appears at the location of the photon sphere of the BH. These findings are consistent with the radius of the photon sphere of the BH, which is calculated in geometrical optics. Our study contributes to a better understanding of the analytical studies of holographic theory, which can be used to evaluate different types of BHs for a fixed wave source and optical system.\\

\end{abstract}
\date{\today}
\maketitle

\section{Introduction}

Since there is a thrust among the research community for the
unification of gravity and the standard model. One of the primary
goals of the analysis of the holographic principle or Anti-de-Sitter
(AdS)/conformal field theory (CFT) correspondence \cite{1,2,3} is to
find what type of quantum field theories (QFTs) demonstrates their
dual gravity conformal. During the last two decades, it has been
imagined that duality can be used to define realistic systems such
as the physics of condensed matter \cite{4,5,6}. The dual pair is
the most prominent example between the type $IIB$ string theory on
the AdS$_{5}\times$S$^{5}$ and the maximally super-symmetric gauge
theory in four dimensions, N$=4$ leads to super Yang-Mills \cite{7}
theory, moves towards the new research possibilities on strong
coupling field theory. Subsequently, the holographic principle of
gravity attains a significant position among the various fields of
physics because it is not only used to indirectly test the
correspondence relates to quantum physics but also provides the
solutions to some problems faced by strong coupling systems
associated with the high energy collective excitations. The
innovative concept of the relation between strong coupling field
theory and Einstein's gravity has been analyzed in various domains,
like as AdS/Quantum Chronodynamics (AdS/QCD) an application of
AdS/CFT duality \cite{8}, quantum phase transition, chiral phase
transition, QCD vacuum and higher-dimensional quantum gravity
systems under the AdS/CFT correspondence \cite{9,10,11,12,japan2}.

Additionally, the usage of AdS/CFT correspondence in the study of
condensed matter physics has drawn a lot of interest \cite{13},
particularly in the areas of superfluidity, superconductivity, Fermi
and non-Fermi liquids, dynamics of BHs, providing an entirely novel
viewpoint on the physics of extreme temperatures super-conducting
materials \cite{14,15,16,17}. Further, the application of AdS/CFT in
quantum information provides us with a significant result on
multi-body systems, like as holographic entanglement entropy
\cite{18}, mutual information \cite{19}, entanglement of
purification \cite{20}, holographic complexity and shooting null
geodesics into holographic space-times \cite{21,22,japan1}.
Meanwhile, some other types of holographic correspondences including
the dS/CFT and Kerr/CFT correspondence were further investigated
under the direction of the analog of dS/CFT \cite{23,24}. Kaku et
al. \cite{japan3} proposed a method to create a star orbiting in an
asymptotically AdS space-time using the AdS/CFT correspondence and
demonstrate an appropriate source in the quantum field theory which
is defined on a $2$-sphere, where the localized star gradually
appears in the dual asymptotically AdS geometry. The theories have
undergone extensive development and have produced many outstanding
results. But it is challenging to put these brilliant theories to
the test in an experiment. More evidence should be acquired to
confirm the reliability of AdS/CFT proposals.

A BH is a fascinating and interesting prediction of Einstein's theory of GR, explaining the dynamics of space-time regions that have experienced gravitational collapse. Recently, the Laser-Interferometer Gravitational Wave-Observatory (LIGO) experiments found the gravitational waves emission from the merger of two BHs, providing strong evidence to prove the confirmation of BHs in compact binaries \cite{25}. The Event Horizon Telescope (EHT) recently publicized the first image of the super-massive BH at the heart of the M$87^{\star}$ giant elliptical galaxy. It put a new spirit among the researchers because it resolved many mysteries about the BH dynamics and its related concepts. The physical appearance of captured BH reports a compact asymmetric ring-like shape, which depicted a bright ring-shaped lump of radiation surrounding a circular dark silhouette, which is the so-called BH shadow. The appearance of dark shadow due to the light rays passing near the BH are absorbed by a BH, which casts a dark shadow in the observers's sky \cite{26,27,28,29,30,31}.

Further, the EHT revealed the corresponding linear polarimetric shadow of the M$87^{\star}$, which is crucial to understanding the emission of matter jets from its core. The accreting of matter around the BH carries significant information about the geometry of the magnetized field responsible for the accelerating emission and found the accretion disks around the M$87^{\star}$ \cite{32}. A BH gives us a constant space-time geometry, which is illuminated by some external sources of the optical accretion material, leading a BH with different structures and emitting a beam of different colors. This makes it feasible to analyze the observational characteristics of BH shadows surrounded by different accretion flow models. In the mechanism of different theories, the study of BH shadow and its observational appearance along with other significant properties have also been analyzed in \cite{33,34,35} and many other associated works exist in literature. Nowadays, a number of researchers have devoted themselves to exploring the physics of BH images and trying to achieve remarkable results, which told a comprehensive story of BH in our Universe.

In this scenario, we concluded that the shadow of BH provides us with significant information about the geometric structure of space-time and explores some interesting features of different gravity models more deeply. However, the present research on BH shadow is based on the null geodesics geometrical optics method. In \cite{36,37}, authors analyzed the holographic image of the AdS BH in bulk and investigated a material correspondence to an AdS space-time, when the scalar operator wave emitted by the source at a finite temperature in the boundary of AdS enters the bulk and then generates the propagations in the bulk. They observed that the Einstein ring can be precisely viewed through the holographic image, and the radius of the photon ring is consistent with the BH photon sphere, which is determined through geometric optics, leading to the existence of dual BH. Using this method, the authors in \cite{38} analyzed the geometry of the Einstein ring for the lensed response function of the complex scalar field potential as the wave propagates in the charged AdS BH within the framework of AdS/CFT. Further, in the context of Maxwell and charged scalar fields, the influence of the charged scalar condensate is analyzed on the photon ring image, where the asymptotic AdS BH image is dual to a superconductor \cite{39}.

During the last two decades, numerous efforts have been made to modify the theory of GR, and in particular try to develop a theory with massive gravitons, the so-called MG. The attraction of MG is that it describes the cosmic expansion of our Universe without invoking the concept of cosmological constant. The influence of putting massive gravitons significantly modified the GR by reducing it to a large scale, which permits the universe to accelerate, although its findings at lower scales are like those in GR. The formulation of MG was initially introduced by Fierz and Pauli \cite{40}, but unfortunately, this theory did not approach the GR limits in the massless framework. Further, another problem was found in this theory leads to ghost instability \cite{41,42}, up to now, a number of efforts have been made to resolve the problem of ghost instability, leading to further modified the MG, namely ghost-free MG, which was introduced in \cite{43,44}. In this scenario, the thermodynamic properties of BHs are analyzed in \cite{45}. The $P$-$V$ criticality and extended phase transition of charged AdS BHs are analyzed in ghost-free MG \cite{46}, in which the cosmological constant plays a role of the dynamical pressure in the BH framework. In \cite{47}, authors analyzed the influence of massive graviton on the holographic thermalization mechanism. They further adopted the two-point correlation formulation at the same time to determine the thermalization mechanism in dual field theory and observed the effect of graviton mass parameter in the framework of AdS/CFT correspondence. Further, in the context of MG, the accretion of matter and shadow of BHs with Lorentz symmetry are studied in \cite{48}.

Considering the AdS/CFT framework, we explicitly demonstrated that
holographic Einstein images of the dual BH from a given response
function on the side of the AdS boundary, where the response
function is generated through some external sources lie far away
from the response function on the AdS boundary. For instance, we
consider as an example $(2+1)$-dimensional boundary CFT on the
$2$-sphere $\mathcal{S}^{2}$ at a finite temperature in the global
AdS$_{4}$ space-time. An oscillatory Gaussian source
$\mathcal{J}_{\mathcal{O}}$ on one side of the AdS boundary, and
scalar waves generated by the source can propagate in the bulk and
reach the other side of the AdS boundary, the corresponding lens
responses will be generated, i.e., $\langle O\rangle$ \cite{zeng3}.
The working principle of this imaging system is presented in Fig.
\textbf{1}.
\begin{figure}[H]\centering
\includegraphics[width=18cm,height=6.4cm]{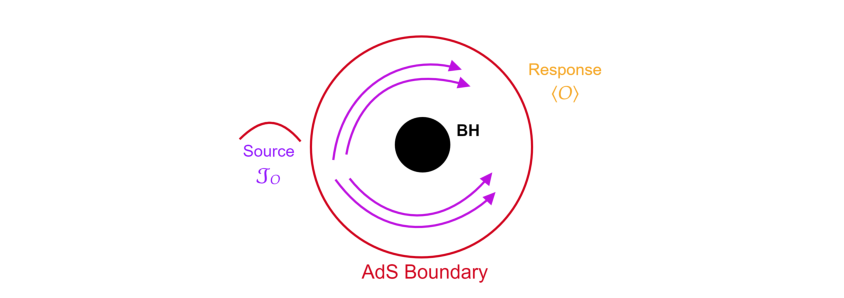} \caption{The constructed schematic diagram for imaging a dual BH, in which Gaussian Source $\mathcal{J}_{\mathcal{O}}$ lies on the AdS boundary and its related response $\langle O\rangle$ lies far away on the same boundary.}
\end{figure}
According to this view, the literature persuades us to investigate the holographic image of MG AdS BH more comprehensively. The main purpose of this work is to use the wave optics method and follow the ideas proposed in \cite{36,37,38} with the AdS/CFT correspondence. We investigate the behavior of the lensed response function for different values of model parameters in the spherically symmetric AdS BH and analyze the possible influence of each parameter on the resulting Einstein ring structure, which may provide a strong signal for the confirmation of its gravity dual. The remainder of the present paper is outlined as follows. In section \textbf{II}, we briefly define the basic formulation of MG AdS BH and holographic setup and extract the corresponding lensed response function. With the help of the optical system, consisting of a convex lens and spherical screen, we investigate the Einstein ring formation for our considering framework and compare our derived results with the optical approximation in section \textbf{III}. In the same section, we also depict the profiles of lensed response brightness and the basic formulation of the ingoing angle of the photon. The last section is devoted to the conclusion.

\section{Holographic Setup and Ring Formation of AdS Black Hole in Massive Gravity}

In the present work, we consider the action of massive gravity, which is written as follows \cite{45,49}
\begin{equation}\label{1}
\mathcal{S}=\frac{1}{16 \pi G}\int
d^4{x\sqrt{-g}}[R+\frac{6}{l^{2}}+m^{2}\sum_{i=1}^4c_{i}\mathcal{H}_{i}(g_{\alpha\beta},f_{\alpha\beta})],
\end{equation}
where $R$ is the scalar curvature, $l$ is the radius of the AdS space-time, $m$ represents the graviton mass parameter, $c_{i}$ are constants and $\mathcal{H}_{i}$ are symmetric polynomials of the eigenvalues of the matrix $\mathcal{K}^{\alpha}_{\beta}\equiv\sqrt{g^{\alpha\mu}f_{\mu\beta}}$ defined as
\begin{eqnarray}\nonumber
\mathcal{H}_{1}&=&[\mathcal{K}],\quad
\mathcal{H}_{2}=[\mathcal{K}]^{2}-[\mathcal{K}^{2}],\quad
\mathcal{H}_{3}=[\mathcal{K}]^{3}-3[\mathcal{K}][\mathcal{K}^{2}]+2[\mathcal{K}^{3}],
\\\label{2}\mathcal{H}_{4}&=&[\mathcal{K}]^{4}-6[\mathcal{K}^{2}][\mathcal{K}]^{2}+
8[\mathcal{K}^{3}][\mathcal{K}]+3[\mathcal{K}^{2}]^{2}-6[\mathcal{K}^{4}].
\end{eqnarray}
The square root in $\mathcal{K}$ means $(\sqrt{A})^{\alpha}_{\mu} (\sqrt{A})^{\mu}_{\beta}=A^{\alpha}_{\beta}$ and $[\mathcal{K}]$ represents the trace $\mathcal{K}^{\alpha}_{\alpha}=\sqrt{g^{\alpha\mu}f_{\mu\alpha}}$. Finally, $f_{\alpha\beta}$ is a fixed symmetric tensor called the reference metric, and the gauge-fixed ansatz for the reference metric is defined as
\begin{equation}\label{3}
f_{\alpha\beta}=\text{diag}(0,0,c_{0}^{2},c_{0}^{2}\sin^{2}\theta).
\end{equation}
Then, one can find the static and spherically symmetric BH solution in the following form as \cite{49}
\begin{equation}\label{4}
ds^{2}=-B(r)dt^{2}+\frac{dr^{2}}{B(r)}+r^{2}d\Omega^{2},
\end{equation}
with
\begin{equation}\label{5}
B(r)=1-\frac{2M}{r}+\frac{r^{2}}{l^{2}}+\frac{c_{0}c_{1}m^{2}}{2}r+c_{0}^{2}c_{2}m^{2},
\end{equation}
in which $M$ represents the mass of the BH,
$d\Omega^{2}=d\theta^{2}+\sin^{2}\theta d\phi^{2}$, $c_{0},~c_{1}$
and $c_{2}$ are the constant parameters associated with graviton
mass. Here, we fix $c_{0}=c_{1}=1$ and $c_{2}=-1/2$, in order to
obtain the thermodynamical stable background \cite{45,49}. Now, we
concentrate on the spherical metric ansatz with a new definition as
$u=1/r$ and $B(r)=u^{-2}B(u)$. Based on these assumptions, one can
write the new coordinate as $(t,u,\theta,\phi)$, which is used to
rewrite the metric function which is defined in Eq. (\ref{4}) with
the following ansatz
\begin{equation}\label{6}
ds^{2}=\frac{1}{u^{2}}[-B(u)dt^{2}+\frac{du^{2}}{B(u)}+d\Omega^{2}].
\end{equation}
The value of $u=\infty$ leads to space-time singularity, while the
CFT boundary corresponds to $u=0$, leading to the existence of a
dual quantum system. In addition, the temperature of the boundary system of the BH is defined by the Hawking temperature, which is given by the relation $T=\frac{1}{4\pi}B'(u_{e})$ (where $u_{e}=u$ represents the event horizon of the BH). Next, we consider the
Klein-Gordon equation which is used to determine the dynamics of a
massless scalar field as \cite{37}
\begin{equation}\label{7}
D_{a}D^{a}\Psi=0.
\end{equation}
Now, we are going to define the ingoing Eddington coordinate to solve the Klein-Gordon equation accurately, which provides insights into the nature of BHs and preserves the smoothness of physical quantities at the event horizon. Mathematically, these coordinates can be written as \cite{38}
\begin{eqnarray}\label{8}
v=t+u_{\star}=t-\int\frac{du}{B(u)},
\end{eqnarray}
leads to the metric function can be redefined as
\begin{equation}\label{9}
ds^{2}=\frac{1}{u^{2}}[-B(u)dv^{2}-2dudv+d\Omega^{2}].
\end{equation}
Near the AdS boundary, the asymptotic solution of the scalar field becomes
\begin{equation}\label{10}
\Psi(v,u,\theta,\phi)=\mathcal{J}_{\mathcal{O}}(v,\theta,\phi)+u\partial_{v}\mathcal{J}_{\mathcal{O}}(v,\theta,\phi)+
\frac{1}{2}u^{2}D^{2}_{\mathcal{S}}\mathcal{J}_{\mathcal{O}}(v,\theta,\phi)+\langle
O\rangle u^{3}+\mathcal{O}(u^{4}),
\end{equation}
where $D^{2}_{\mathcal{S}}$ denotes the scalar Laplacian on unit $\mathcal{S}^{2}$. Based on the AdS/CFT framework, the terms $\mathcal{J}_{\mathcal{O}}$ and $\langle O\rangle$ leading to an external scalar source and the corresponding response function in dual CFT, respectively \cite{50}. In the present work, we consider the axis-symmetric and mono-chromatic Gaussian wave packet at the south pole ($\theta=\pi$) of the AdS boundary. In this scenario, we have
\begin{eqnarray}\label{11}
\mathcal{J}_{\mathcal{O}}(v,\theta)=e^{-i\omega
v}\exp[-(\pi-\theta)^{2}/2\tau^{2}]/2\pi\tau^{2}= e^{-i\omega
v}\sum_{l=0}^{\infty} C_{l0}X_{lo}(\theta),
\end{eqnarray}
where $\tau$ is the width of the wave, which is produced by the Gaussian source and $X_{lo}(\theta)$ is the spherical harmonics function. We ignore the Gaussian tail safely due to its smallest value and hence we only suppose the case $\tau\ll\pi$. Further, the coefficients of $X_{lo}(\theta)$ can be calculated as
\begin{equation}\label{12}
C_{l0}=(-1)^{l}((l+1/2)/2\pi)^{\frac{1}{2}}\exp\bigg[-\frac{(l+1/2)^{2}\tau^{2}}{2}\bigg].
\end{equation}
Now considering the symmetry of space-time, the scalar field further decomposed $\Psi(v,u,\theta,\phi)$ as
\begin{equation}\label{13}
\Psi(v,u,\theta,\phi)=e^{-i\omega
v}\sum_{l=0}^{\infty}\sum_{n=-l}^{l}C_{l0}U_{l}(u)X_{ln}(\theta,\phi),
\end{equation}
and the response function $\langle O\rangle$, simultaneously written as
\begin{equation}\label{14}
\langle O\rangle=e^{-i\omega v}\sum_{l=0}^{\infty}C_{l0}\langle
O\rangle_{l} X_{lo}(\theta).
\end{equation}
With the help of Eq. (\ref{13}), we obtain $U_{l}$, satisfying the equation of motion as
\begin{equation}\label{15}
u^{2}B(u)U''_{l}+[u^{2}B'(u)-2uB(u)+2i\omega u^{2}]U'_{l}+[-2i\omega
u-l(l+1)u^{2}]U_{l}=0,
\end{equation}
where the asymptotic behavior of $U_{l}$ can be defined in the following form
\begin{equation}\label{16}
\lim\limits_{u \to 0}U_{l}=1-i\omega
u+\frac{u^{2}}{2}(-l(1+l))+\langle
O\rangle_{l}u^{3}+\mathcal{O}(u^{4}).
\end{equation}
From Eq. (\ref{13}), clearly there are two boundary conditions for the function $U_{l}$. One of them is horizon boundary condition, at the event horizon $u=u_{e}$, has the following form
\begin{equation}\label{17}
U'_{l}[B'(u_{e})u_{e}^{2}+2i\omega u_{e}^{2}]-[2i\omega
u_{e}+l(l+1)u_{e}^{2}]U_{l}=0.
\end{equation}
The other is at space infinity point $u=0$, and the source behaves
as the scalar field, i.e., we have $U_{l}(0)=1$, which is derived
through Eqs. (\ref{11}) and (\ref{13}). Further, we employ the
pseudo-spectral method and derived the desired numerical results for
$U_{l}$ and extract $\langle O\rangle_{l}$ \cite{36,37}. Using the
value of $\langle O\rangle_{l}$, one can obtain the value of the
total response function through Eq. (\ref{14}). We exhibit a profile
of the total response in Figs. \textbf{2} to \textbf{4}, in which
the optical appearance arises from the diffraction of the scalar
field of the BH. We plot Fig. \textbf{2}, which shows the behavior
of amplitude by varying the graviton mass parameter $m$ and choosing
some specific values of other involved parameters as $M=1,~u_{e}=1$
and angular frequency $\omega=80$. Similarly, Fig. \textbf{3} shows
the oscillation period of the wave amplitude for different values of
$\omega$ and choosing some specific values of other involved
parameters as $M=1,~u_{e}=1$ and $m=1$. Figure. \textbf{4} also
depicted the behavior of the amplitude by varying the temperature
$T$ of the boundary system and choosing some specific values of
other involved parameters as $M=1,~m=0.6 $ and angular frequency
$\omega=80$. From Fig. \textbf{2}. it can be seen clearly; the
amplitude of the total response function shows the increasing
behavior with the increasing values of $m$. In Fig. \textbf{3}, the
oscillation period of the wave amplitude is maximum when $\omega=60$
and decreasing smoothly when $\omega=20$ and $\omega=40$. The
amplitude of the total response function significantly varies with
temperature, for instance, when $T=0.776$, the amplitude reaches a
peak position and moves down at $T=0.474$ and $0.333$ nicely, see
Fig. \textbf{4}. This means that the amplitude of the total response
function increases with the increasing values of $T$. Hence, the
observed images of the transformed response function may help to
reflect the significant features of space-time geometry.
\begin{figure}[H]\centering
\epsfig{file=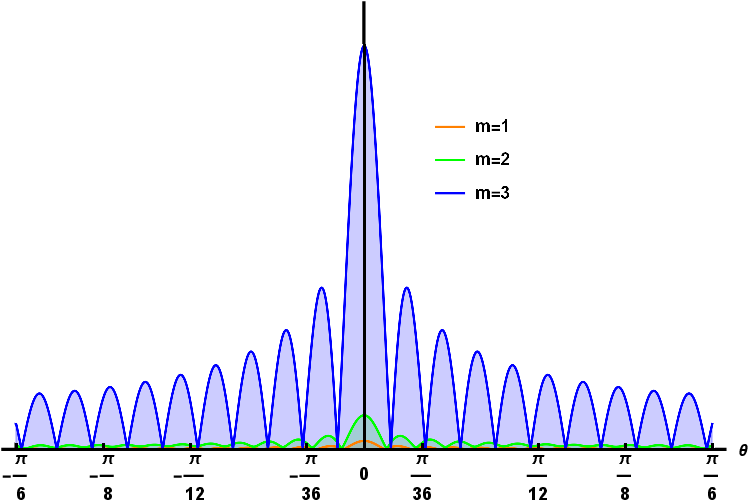, width=.5\linewidth} \caption{The absolute
amplitude of $\langle O\rangle$ around the north pole for various
values of $m$ with $u_{e}=1$ and $\omega=80$.}
\end{figure}
\begin{figure}[H]\centering
\epsfig{file=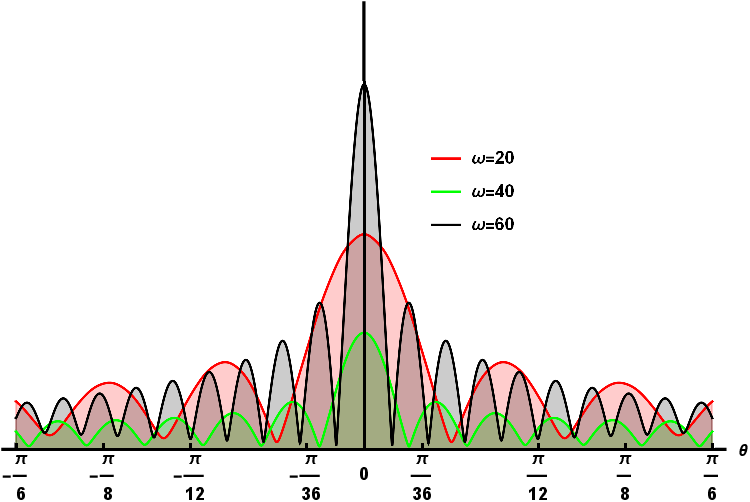, width=.5\linewidth} \caption{The absolute
amplitude of $\langle O\rangle$ around the north pole for various
values of $\omega$ with $u_{e}=m=1$.}
\end{figure}
\begin{figure}[H]\centering
\epsfig{file=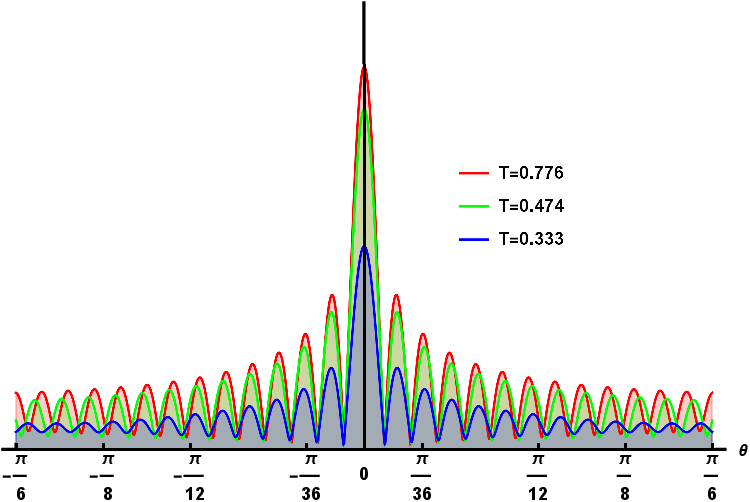, width=.5\linewidth} \caption{The absolute amplitude of $\langle O\rangle$ around the north pole for various values of $T$ with $m=0.6$ and $\omega=80$. Here, the red, green, and blue colors correspond to $u_{e}=0.6,~0.8$, and $1$, respectively.}
\end{figure}

\section{Holographic Rings Formation}

As we observed above, the interference pattern was generated due to
the diffraction of the scalar wave optics by the BH, and we consider
all amplitudes from different sides or angles without any
differentiations. It is a natural phenomenon; our eyes should
distinguish the light patterns from different angles or observe one
color from all sides. For a comprehensive view of the BH, we need to
investigate the response function through an optical framework with
a convex lens, in which the ingoing angle provides us significant
information for observation. In this perspective, we set a
telescope, which is used to analyze the frequency domain of the
responses from different angles at the boundary, as shown in Fig.
\textbf{5}. The convex lens is used to transform the plane wave into
the spherical wave and the received image from the transmitted wave
is depicted on the screen.
\begin{figure}[H]\centering
\includegraphics[width=15cm,height=6.8cm]{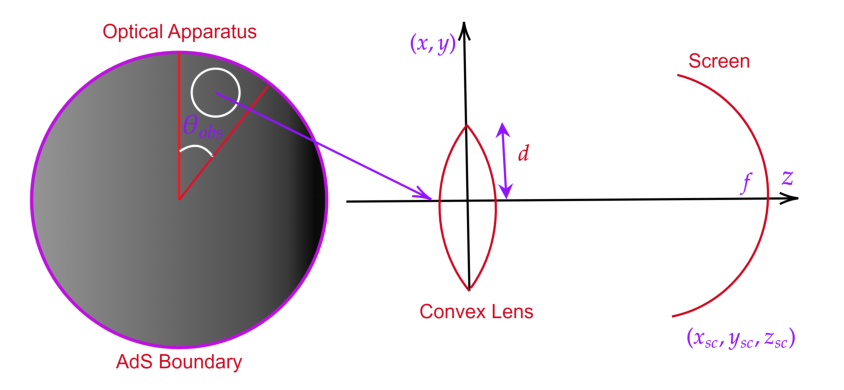} \caption{The structure of image formation system. Where the observational region is surrounded by the white circle on the AdS boundary.}
\end{figure}
We consider an observer to be located at the center of the small circle, where ($\theta,~\vartheta)=(\theta_{obs},~0$) on the AdS boundary and observer looks up into the AdS bulk at this point. We introduce a new polar coordinates as ($\theta',~\vartheta'$), satisfying
\begin{equation}\label{18}
\sin\theta'\cos\vartheta'=e^{i\theta_{obs}}(\sin\theta\cos\vartheta+i\cos\theta),
\end{equation}
which provide guarantees that $\theta'=0,~\vartheta'=0$ corresponds to the observational point. For a virtual optical system, we suppose a cartesian coordinate system ($x,y,z$) with $(x,y)=(\theta'\cos\vartheta',~\theta'\sin\vartheta')$ at the boundary of the observational point. We adjust the convex lens in the two-dimensional $(x,y)$ plane, in which the focal length of the lens and corresponding radius are denoted by $f$ and $d$, respectively. Further, we define the coordinates name on the spherical screen as $(x,y,z)=(x_{sc},~y_{sc},~z_{sc})$ satisfying $x^{2}_{sc}+~y^{2}_{sc}+~z^{2}_{sc}=f^{2}$ \cite{36,37,39}. Consider a wave $\hat{\Psi}(\widehat{x})$ having a frequency $\omega$, which is obtained through the convex lens, and the transmitted wave $\hat{\Psi}_{T}(\widehat{x})$, is defined as
\begin{equation}\label{19}
\hat{\Psi}_{T}(\widehat{x})=e^{-i\omega\frac{|\widehat{x}|^{2}}{2f}}\hat{\Psi}(\widehat{x}).
\end{equation}
Now, the wave function imaging on the screen becomes
\begin{eqnarray}\label{20}
\hat{\Psi}_{sc}(\widehat{x}_{sc})=\int_{|\widehat{x}|\leq
d}d^{2}x\hat{\Psi}_{T}(\widehat{x})e^{i\omega \mathcal{D}}\propto
\int_{|\widehat{x}|\leq
d}d^{2}x\hat{\Psi}(\widehat{x})e^{-i\frac{\omega}{f}\widehat{x}.\widehat{x}_{sc}}=\int
d^{2}x\hat{\Psi}(\widehat{x})\eta(\widehat{x})e^{-i\frac{\omega}{f}\widehat{x}.\widehat{x}_{sc}},
\end{eqnarray}
in which $\mathcal{D}$ is the distance from the lens point $(x,y,0)$ to the screen point ($x^{2}_{sc},~y^{2}_{sc},~z^{2}_{sc}$) and $\eta(\widehat{x})$ is the window function, which is defined as
\begin{equation}\label{21}
\eta(\widehat{x})=
    \begin{cases}
     \text{$1,$ \quad $0\leq|\widehat{x}|\leq d$}, \\
     \text{$0$,\quad ~$|\widehat{x}|>d$}.
    \end{cases}
\end{equation}
\begin{figure}[H]
\begin{center}
\subfigure[\tiny][~$m=1,~\theta_{obs}=0$]{\label{a1}\includegraphics[width=3.9cm,height=4.0cm]{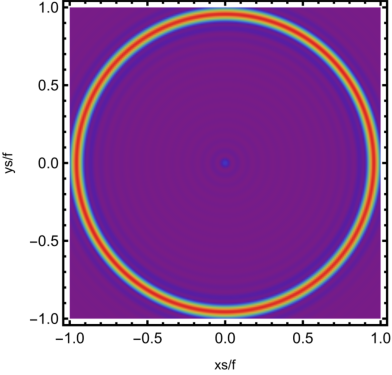}}
\subfigure[\tiny][~$m=1,~\theta_{obs}=\pi/6$]{\label{b1}\includegraphics[width=3.9cm,height=4.0cm]{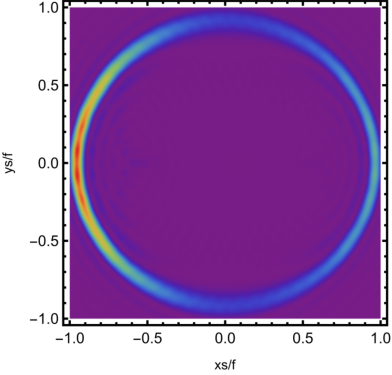}}
\subfigure[\tiny][~$m=1,~\theta_{obs}=\pi/3$]{\label{c1}\includegraphics[width=3.9cm,height=4.0cm]{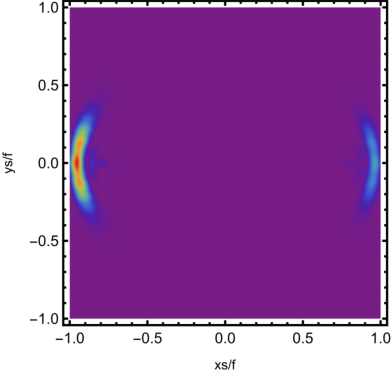}}
\subfigure[\tiny][~$m=1,~\theta_{obs}=\pi/2$]{\label{d1}\includegraphics[width=3.9cm,height=4.0cm]{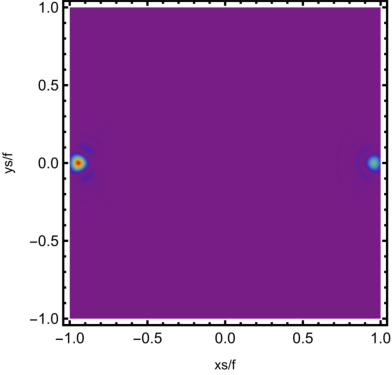}}
\subfigure[\tiny][~$m=2,~\theta_{obs}=0$]{\label{a1}\includegraphics[width=3.9cm,height=4.0cm]{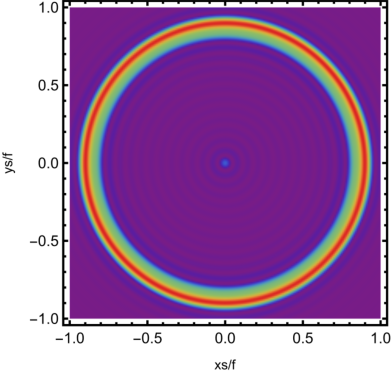}}
\subfigure[\tiny][~$m=2,~\theta_{obs}=\pi/6$]{\label{b1}\includegraphics[width=3.9cm,height=4.0cm]{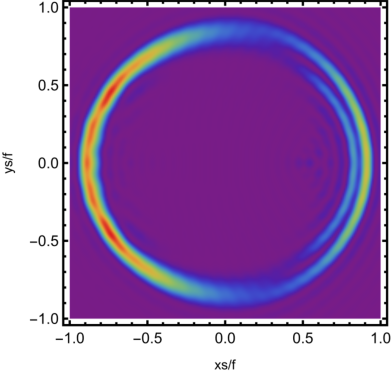}}
\subfigure[\tiny][~$m=2,~\theta_{obs}=\pi/3$]{\label{c1}\includegraphics[width=3.9cm,height=4.0cm]{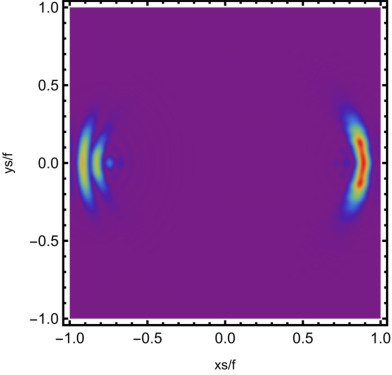}}
\subfigure[\tiny][~$m=2,~\theta_{obs}=\pi/2$]{\label{d1}\includegraphics[width=3.9cm,height=4.0cm]{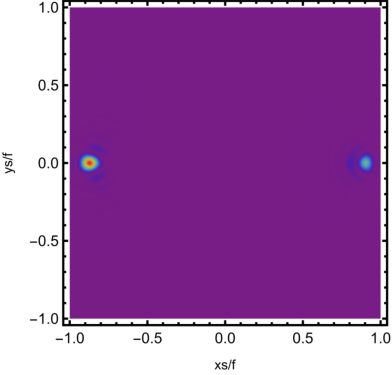}}
\subfigure[\tiny][~$m=3,~\theta_{obs}=0$]{\label{a1}\includegraphics[width=3.9cm,height=4.0cm]{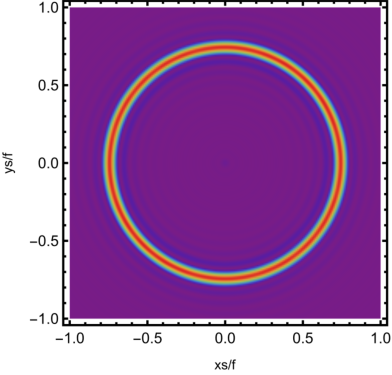}}
\subfigure[\tiny][~$m=3,~\theta_{obs}=\pi/6$]{\label{b1}\includegraphics[width=3.9cm,height=4.0cm]{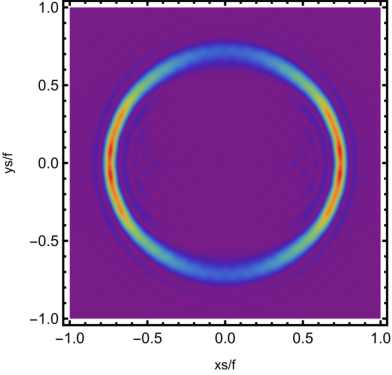}}
\subfigure[\tiny][~$m=3,~\theta_{obs}=\pi/3$]{\label{c1}\includegraphics[width=3.9cm,height=4.0cm]{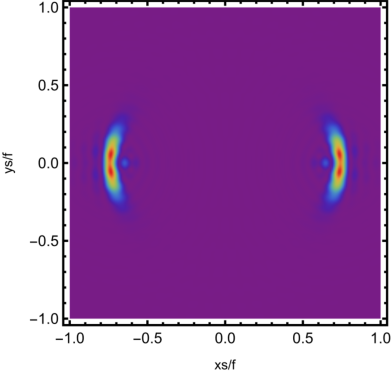}}
\subfigure[\tiny][~$m=3,~\theta_{obs}=\pi/2$]{\label{d1}\includegraphics[width=3.9cm,height=4.0cm]{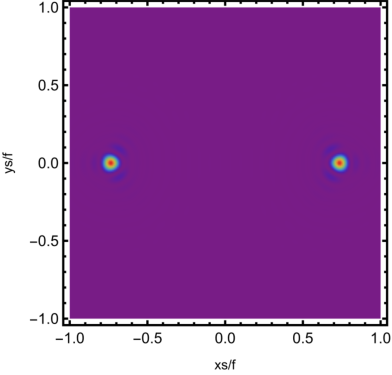}}
\subfigure[\tiny][~$m=4,~\theta_{obs}=0$]{\label{a1}\includegraphics[width=3.9cm,height=4.0cm]{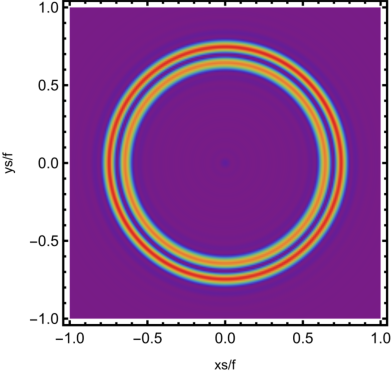}}
\subfigure[\tiny][~$m=4,~\theta_{obs}=\pi/6$]{\label{b1}\includegraphics[width=3.9cm,height=4.0cm]{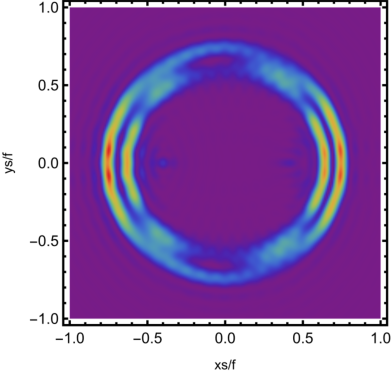}}
\subfigure[\tiny][~$m=4,~\theta_{obs}=\pi/3$]{\label{c1}\includegraphics[width=3.9cm,height=4.0cm]{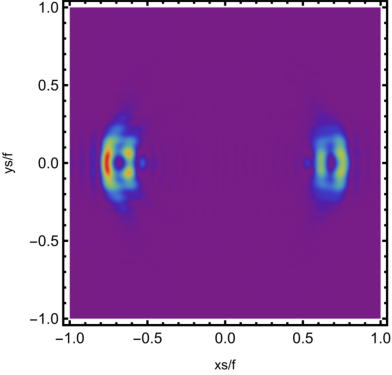}}
\subfigure[\tiny][~$m=4,~\theta_{obs}=\pi/2$]{\label{d1}\includegraphics[width=3.9cm,height=4.0cm]{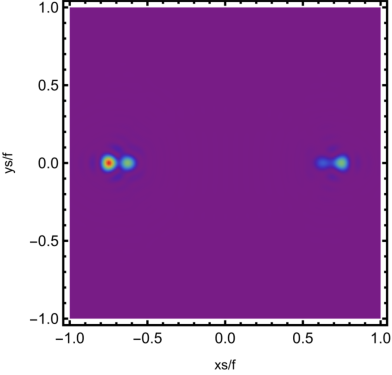}}
\caption{The two-dimensional density plots of the lensed response on the screen for various $m$ with $u_{e}=1$ and $\omega=80$.}
 \end{center}
\end{figure}
From Eq. (\ref{20}), one can see clearly that the observed wave on
the screen is associated with the incident wave through the Fourier
transformation. We will capture the profiles of the dual BH on the
Observer's screen through Eq. (\ref{20}). We will observe different
image profiles of the AdS boundary, the holographic Einstein images
for various values of $m$, and some fixed values of other involved
parameters are depicted in Fig. \textbf{6}. At $\theta=0^{o}$, the
observer is located at the north pole of the AdS boundary, here a
series of axis-symmetric concentric circular rings appear in the
image, and we find only the brightest rings at the north pole as
exhibited in the left column of Fig. \textbf{6}. Further, the values
of BH graviton mass parameter $m$ increase from top to bottom, and
we observe that in all panels the luminosity of the circular-shaped
rings is almost the same.

When we consider $\theta=30^{o}$ (see second column, from top to bottom), the value of BH graviton mass parameter $m$ increases, and the optical appearance of the bright ring changed into a luminosity-deformed ring instead of a continuously strict axis-symmetric ring. Further, as the values of $m$ increase, the ring picks up the extra brightness from the center of the left and right sides and then moves towards the darker region, leading to the continuous vanishing of the shining part of the ring and we observe there will be a low resolution in this region when $m=4$. Further, when we fix the observer's location at $\theta=60^{o}$, one can see that t only bright arcs have appeared, and bright arcs exist in the middle of the screen and show small brightness as $m$ increases. Consequently, when $\theta=90^{o}$, there exist two tiny bright spots in which one of them has small brightness as compared to the other, and when $m=3$, these two spots show almost the same brightness. After that when $m=4$, these two spots are further divided into four spots and the right side shows smaller brightness as compared to the left one. From the above discussions, we conclude that the parameter $m$ has a significant influence on the position of the holographic Einstein image, which is used to investigate some structural properties of the BH.

Now, we are going to investigate the effect of horizon temperature
on the profiles of the lensed response as shown in Fig. \textbf{7},
which is observed for some fixed values of the involved parameters
such as $\theta_{obs}=0^{o},~m=1$ and $\omega=80$. We depicted the
observational image of the dual BH for increasing values of the
horizon. When $u_{e}=0.1$, there is only one bright spot in the
center of the screen, and here $T=24.7088$. With the increasing
value of the horizon, such as $u_{e}=1.1$, we observe a series of
axis-symmetric bright rings in the image, here $T=0.30943$, and one
particular bright ring lies far away from the central region. At
$u_{e}=20.1$, we have $T=0.044339$, and observe that the small
brightest ring lies near the center. Further, when $u_{e}=40.1$, we
observe that $T=0.041921$, we find the bright smallest ring in the
center and come closer to the center as compared to the previous
one.
\begin{figure}[H]
\begin{center}
\subfigure[\tiny][~$T=24.7088$]{\label{a1}\includegraphics[width=3.9cm,height=4.0cm]{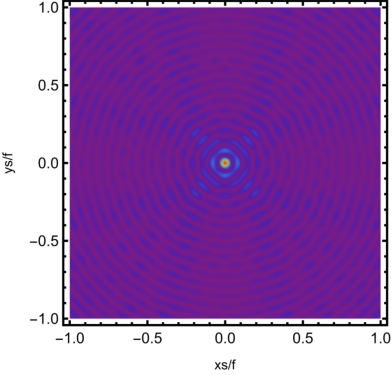}}
\subfigure[\tiny][~$T=0.30943$]{\label{b1}\includegraphics[width=3.9cm,height=4.0cm]{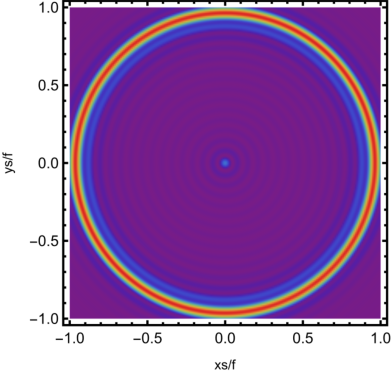}}
\subfigure[\tiny][~$T=0.044339$]{\label{c1}\includegraphics[width=3.9cm,height=4.0cm]{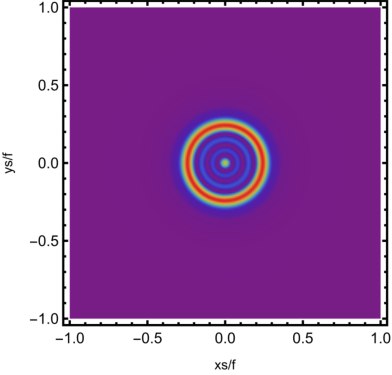}}
\subfigure[\tiny][~$T=0.041921$]{\label{d1}\includegraphics[width=3.9cm,height=4.0cm]{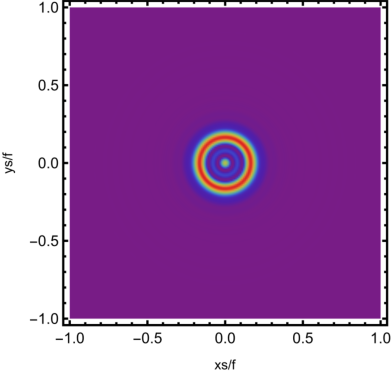}}
\caption{The two-dimensional density plots of the lensed response on
the screen for various $T$ at the observational angle $\theta=0^{o}$
with $m=1$ and $\omega=80$.}
 \end{center}
\end{figure}
\begin{figure}[H]
\begin{center}
\subfigure[\tiny][~$T=24.7088$]{\label{a1}\includegraphics[width=3.9cm,height=4.0cm]{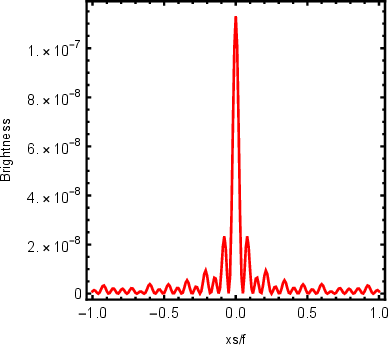}}
\subfigure[\tiny][~$T=0.30943$]{\label{b1}\includegraphics[width=3.9cm,height=4.0cm]{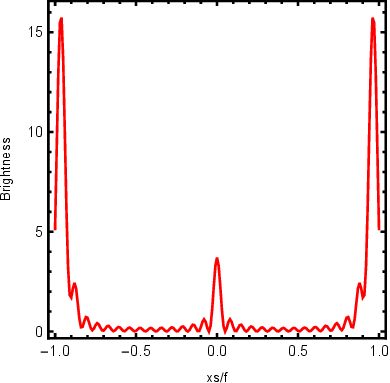}}
\subfigure[\tiny][~$T=0.044339$]{\label{c1}\includegraphics[width=3.9cm,height=4.0cm]{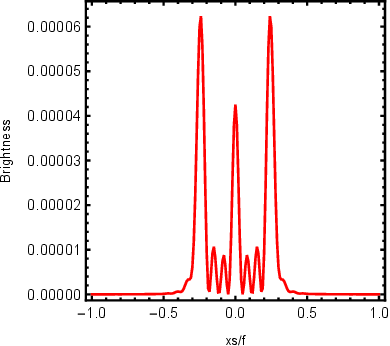}}
\subfigure[\tiny][~$T=0.041921$]{\label{d1}\includegraphics[width=3.9cm,height=4.0cm]{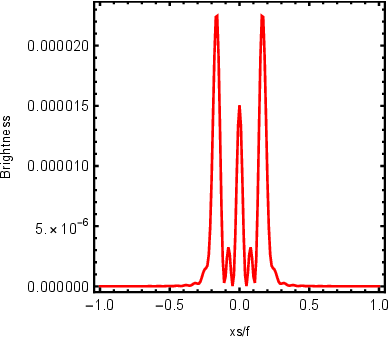}}
\caption{The profiles of the lensed response brightness on the
screen for various $T$ with $m=1$ and $\omega=80$.}
 \end{center}
\end{figure}
For a comprehensive understanding of Fig. \textbf{7}, we also plot
the corresponding brightness profiles in Fig. \textbf{8}
alternatively. In Fig. \textbf{8}(a), the trajectory reaches its
peak value in the center, which corresponds to the bright spot in
Fig. \textbf{7}(a). And when $T=0.30943,~0.044339$ and $0.041921$,
there exist two peaks trajectories in the panels which is
corresponding to a series of axis-symmetric rings of Figs.
\textbf{7}(b), \textbf{7}(c), and \textbf{7}(d), respectively. From
these figures, we concluded that when the temperature of the horizon
is lower, the bright ring lies at the focal point, and higher
values, it gradually moves toward the interior. We further
investigate the influence of the BH graviton mass parameter $m$ on
the profiles of the dual BH. We plot the trajectories of the
brightness for different values of $m$ as shown in Fig. \textbf{9},
where $y$-axis and $x$-axis are showing the intensity and position
of the brightness of the lensed response on the screen,
respectively. As the value of parameter $m$ increases, the
brightness also increases, leading to increasing the luminosity of
the rings. Hence, the holographic profiles of AdS BH can not only
use to analyze the geometrical properties of the BH but also
describe the optical properties of the lens and the wave packet
source significantly.
\begin{figure}[H]
\begin{center}
\subfigure[\tiny][~$m=1$]{\label{a1}\includegraphics[width=3.9cm,height=4.0cm]{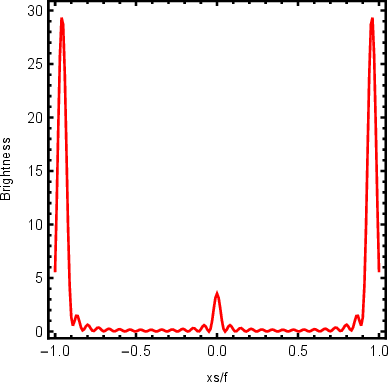}}
\subfigure[\tiny][~$m=2$]{\label{b1}\includegraphics[width=3.9cm,height=4.0cm]{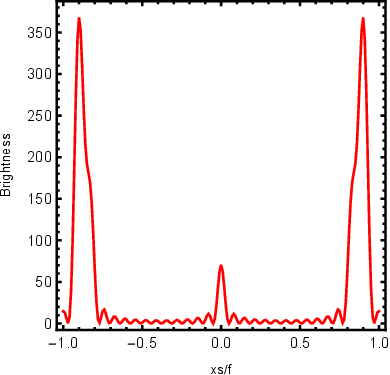}}
\subfigure[\tiny][~$m=3$]{\label{c1}\includegraphics[width=3.9cm,height=4.0cm]{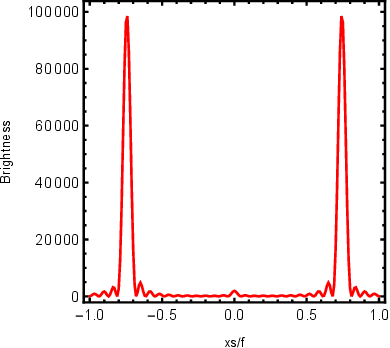}}
\subfigure[\tiny][~$m=4$]{\label{d1}\includegraphics[width=3.9cm,height=4.0cm]{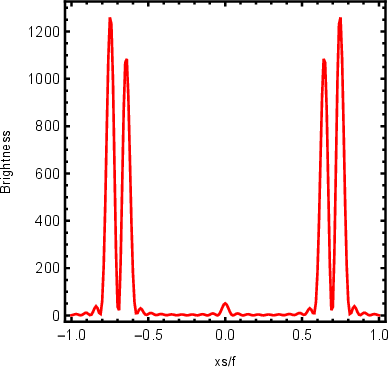}}
\caption{The profiles of the lensed response brightness on the screen for various values of $m$ with $u_{e}=1$ and $\omega=80$.}
 \end{center}
\end{figure}
\begin{figure}[H]
\begin{center}
\subfigure[\tiny][~$\omega=70$]{\label{a1}\includegraphics[width=3.9cm,height=4.0cm]{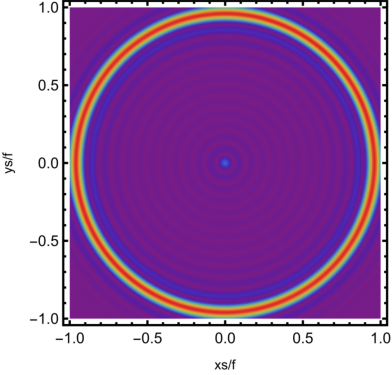}}
\subfigure[\tiny][~$\omega=50$]{\label{b1}\includegraphics[width=3.9cm,height=4.0cm]{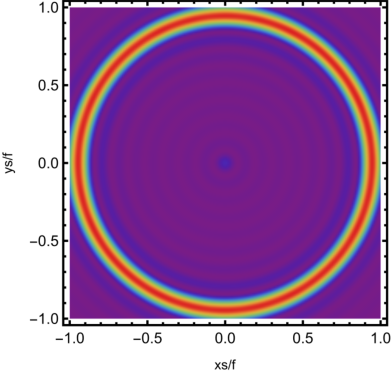}}
\subfigure[\tiny][~$\omega=30$]{\label{c1}\includegraphics[width=3.9cm,height=4.0cm]{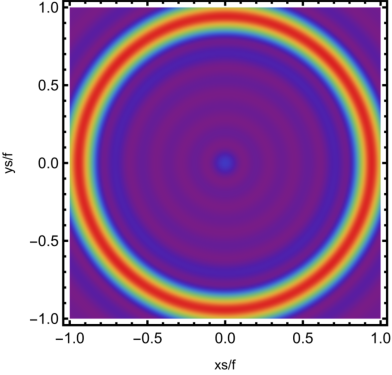}}
\subfigure[\tiny][~$\omega=10$]{\label{d1}\includegraphics[width=3.9cm,height=4.0cm]{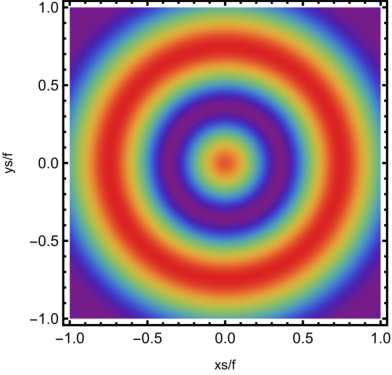}}
\caption{The two-dimensional density plots of the lensed response on the screen for various values of $\omega$ at an observational angle $\theta=0^{o}$ with $m=1$ and $u_{e}=1$.}
 \end{center}
\end{figure}
\begin{figure}[H]
\begin{center}
\subfigure[\tiny][~$\omega=70$]{\label{a1}\includegraphics[width=3.9cm,height=4.0cm]{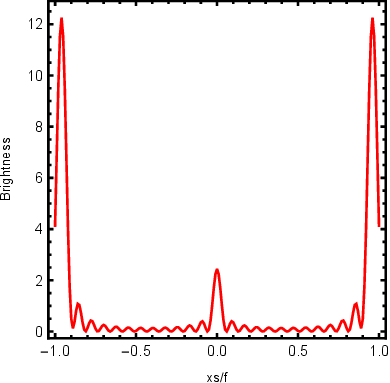}}
\subfigure[\tiny][~$\omega=50$]{\label{b1}\includegraphics[width=3.9cm,height=4.0cm]{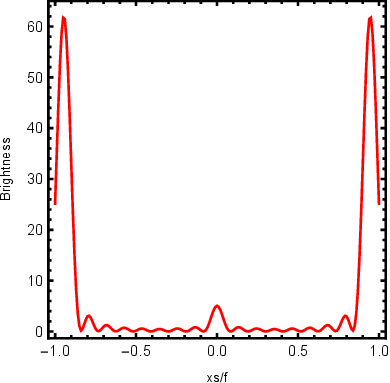}}
\subfigure[\tiny][~$\omega=30$]{\label{c1}\includegraphics[width=3.9cm,height=4.0cm]{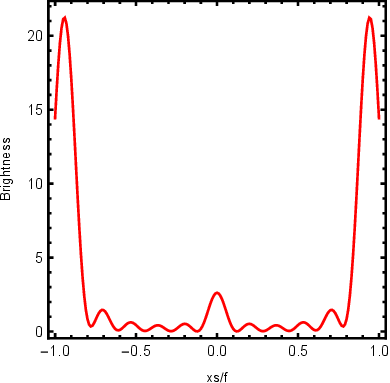}}
\subfigure[\tiny][~$\omega=10$]{\label{d1}\includegraphics[width=3.9cm,height=4.0cm]{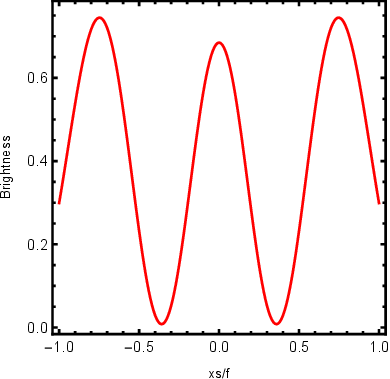}}
\caption{The profiles of the lensed response brightness on the screen for various values of $\omega$ with $m=1$ and $u_{e}=1$.}
 \end{center}
\end{figure}
In addition, the properties of the holographic Einstein picture
under the influence of wave source are depicted in Fig. \textbf{10},
in which we consider $\tau=0.02$ and $d=0.6$ for the convex lens.
One can see, as we increase the value of the frequency, the
resulting ring becomes sharper. This effect makes sense since the
geometric optics approximation framework may capture the image
beautifully in the high-frequency limit. For a comprehensive
understanding of Fig. \textbf{10}, the corresponding profiles of the
lensed response function are also depicted in Fig. \textbf{11}.
Here, one can also see different values of the frequency lead to
changing the lensed response brightness, which is also characterized
by the middle row of the image.

At the position of the photon sphere of the BH, there exists the brightest ring in the image. Next, we will verify this bright ring in the image from the perspective of optical geometry. We described the deflection of light caused by the MG BH i.e., the motions of photons around the BH. In space-time with the metric given in Eq. (\ref{4}), we can describe the ingoing angle of photons from the boundary with their invariants $\tilde{\omega}=B(r)\partial t/\partial\kappa$ and $L=r^{2}\partial\phi/\partial\kappa$ in which $\tilde{\omega}$ is the conserved energy of the photon, $\kappa$ is the affine parameter of photon orbit and $L$ is the angular momentum of the photon. Without loss of generality, we consider a coordinate system to let the photon orbit lying on the equatorial $\theta\equiv\frac{\pi}{2}$. The four-vector $v^{\gamma}\equiv(d/d\kappa)^{\gamma}$ satisfies \cite{38,zeng4}
\begin{equation}\label{30}
-B(r)\bigg(\frac{dt}{d\kappa}\bigg)^{2}+\frac{1}{B(r)}\bigg(\frac{dr}{d\kappa}\bigg)^{2}+r^{2}\sin^{2}
\theta\bigg(\frac{d\phi}{d\kappa}\bigg)^{2}=0,
\end{equation}
or equivalently
\begin{equation}\label{31}
\dot{r}^{2}=\tilde{\omega}^{2}-L^{2}u(r),
\end{equation}
here $u(r)=B(r)/r^{2}$ and $\dot{r}=\partial r/\partial\kappa$. The ingoing angle $\theta_{in}$ with normal vector of boundary $n^{\gamma}\equiv\partial/\partial r^{\gamma}$ should be defined as follows
\begin{eqnarray}\label{32}
\cos\theta_{in}=\frac{g_{\alpha\beta}v^{\alpha}n^{\beta}}{|v|
|n|}\bigg|_{r=\infty}=\sqrt{\frac{\dot{r}^{2}/B}{\dot{r}^{2}/B+L^{2}/r^{2}}}\bigg|_{r=\infty},
\end{eqnarray}
which means that
\begin{eqnarray}\label{33}
\sin\theta^{2}_{in}=1-\cos\theta^{2}_{in}=\frac{L^{2}u(r)}{\dot{r}^{2}+L^{2}u(r)}\bigg|_{r=\infty}=\frac{L^{2}}{\tilde{\omega}^{2}}.
\end{eqnarray}
So, the ingoing angle of photon orbit from the boundary satisfies the following relation
\begin{equation}\label{34}
\sin\theta_{in}=\frac{L}{\tilde{\omega}},
\end{equation}
which is depicted in Fig. \textbf{12}. In particular, when the trajectory of the photon reached the position of the photon sphere, then photons will neither escape from the BH nor fall into the BH and start to move around the BH with constant rotation. Further, we suppose the dominant contribution to the final response function, which is come from the special angular momentum as $L_{s}$ with the light trajectory originating from the south pole on the AdS boundary may enter the circular orbit \cite{36,37,38}, which is determined by the conditions as given below
\begin{eqnarray}\label{29}
\dot{r}=0,\quad \frac{du}{dr}=0.
\end{eqnarray}
\begin{figure}[H]\centering
\includegraphics[width=10cm,height=7.25cm]{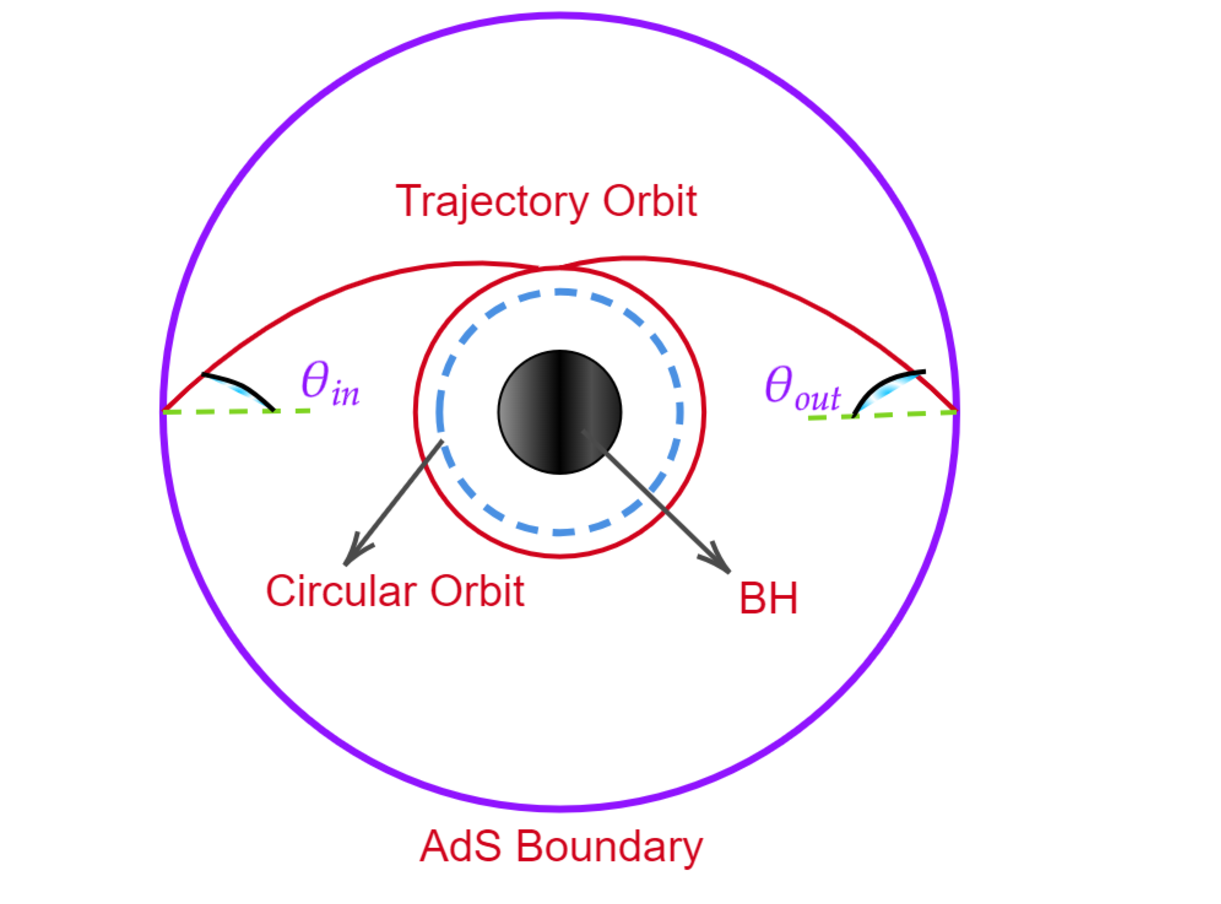} \caption{A schematic diagram of the trajectory orbit of the incident photon revolving around the BH one time.}
\end{figure}
In geometric optics, the angle $\theta_{in}$ gives the angular distance of the image of the incident ray from the zenith if an observer on the AdS boundary looks up into the AdS bulk. If two end points of the geodesic and the center of the BH are in alignment, the observer see a ring image with a radius corresponding to the incident angle $\theta_{in}$ because of axisymmetry \cite{37}. In addition, we expect to see an Einstein ring formed on the screen having the ring radius as follows
\begin{equation}\label{new1}
\sin\theta_{R}=\frac{r_{R}}{f},
\end{equation}
\begin{figure}[H]\centering
\includegraphics[width=10cm,height=7.5cm]{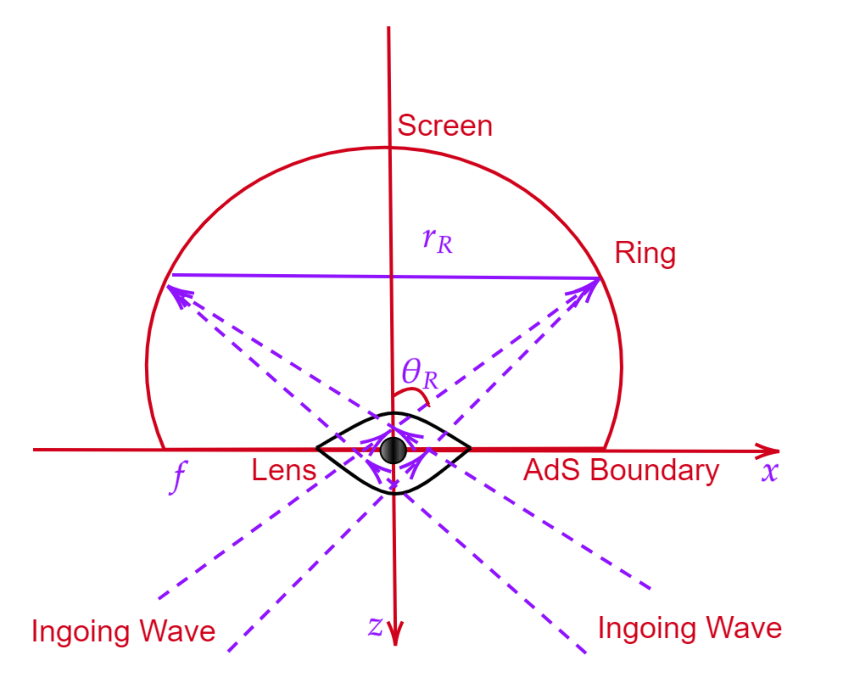} \caption{Schematic diagram expresses the relation between ring radius and ring angle.}
\end{figure}
as shown in Fig. \textbf{13}. According to \cite{37}, when the
angular momentum is sufficiently large, i.e.,
$\sin\theta_{R}=\sin\theta_{in}$, we have following relation
\begin{equation}\label{new55}
\frac{r_{R}}{f}=\frac{L_{s}}{\tilde{\omega}}.
\end{equation}

In fact, both the incident angle of the photon and the angle of the photon ring describe the angle at that the viewer can observe the photon ring, which should be essentially equal, and we will confirm this result from the numerical point of view. Figure \textbf{14} depicted the Einstein ring radius for different values of the parameter $m$, where $r_{R}$ exhibit ring radius in the unit of $f$ as a function of temperature. As expected, one can see that the Einstein
ring radius obtained by our considering wave optics fits well with
that of geometric optics, as the red dots are always located on the blue line or its vicinity.
\begin{figure}[H]
\begin{center}
\subfigure[\tiny][~$m=0.1$]{\label{a1}\includegraphics[width=4.2cm,height=4.3cm]{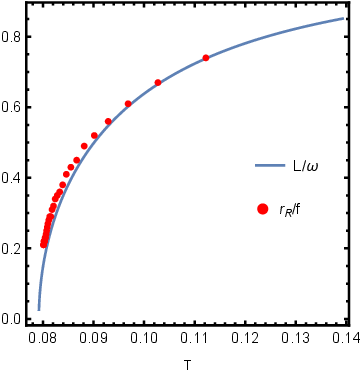}}
\subfigure[\tiny][~$m=0.6$]{\label{b1}\includegraphics[width=4.2cm,height=4.3cm]{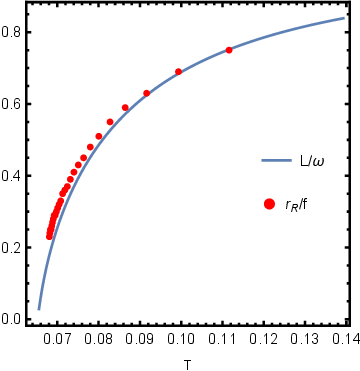}}
\subfigure[\tiny][~$m=1$]{\label{c1}\includegraphics[width=4.2cm,height=4.3cm]{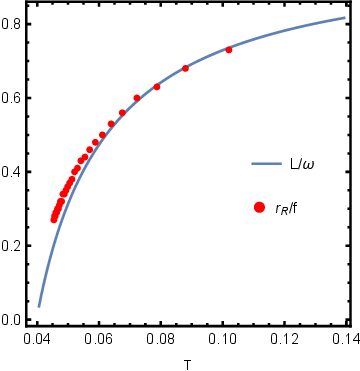}}
\caption{Express the Einstein ring radius as a function of the
temperature for different values of $m$ with $\omega=80$. Where the
discrete red points represent the Einstein ring radius, obtained by
the wave optics, and the blue curve expresses the radius of the
circular orbit varying with temperature, which is obtained from the
geometric optics.}
 \end{center}
\end{figure}

\section{Conclusions}

History has shown that it is impossible to derive physics from astrophysics without understanding astrophysics. The crucial question is not whether current or future EHT results depend on astrophysics but rather how significant the uncertainty due to unknown astrophysics is. The EHT has gone to great lengths to quantify this uncertainty using a wide range of self-consistent numerical simulations and interpreting the interesting phenomenon of BHs dynamics which may accomplish the possible relevant ambiguities.
In this perspective, considering the mechanism of AdS/CFT, we have investigated the holographic images of a MG AdS BH by applying the method of wave optics such as the oscillating Gaussian source produced at the boundary of AdS entering the bulk and propagating in the bulk space-time. After passing the scalar wave through the BH, the resulting profiles show that there always found the diffraction pattern of the total response function at a finite temperature. We observe that the absolute amplitude around the north pole does not closely depend on the space-time geometry i.e., the BH graviton mass parameter $m$ but also significantly varies with the influence of source properties i.e., the width $\tau$ and frequency $\omega$ of the wave.

Further, we derive the local response function through Fourier transformation and depicted the Einstein images of AdS BH with an optical system, which consists of a convex lens and screen. The resulting images show that the radius of the Einstein ring depends on the parameter $m$ leads to variations in temperature or different observational angles of the horizon. To be more specific, when an observer changes the positions, the luminosity of Einstein's ring will change continuously along with light arcs or a bright light spot that appears in the center of the screen. For instance, one can see that when the observational angle increases, the Einstein ring is broken, and only $y$-axis symmetry exists, especially when $\theta_{obs}=\pi/2$, there appear only two points at two symmetrical positions. Further, we also analyzed the brightness of the lensed response function for different values of the model parameters, for example, the brightness of the lensed response function increases with the increasing values of the $m$ and similarly other involved parameters also significantly affected the brightness of the lensed response function.

Moreover, the brightest ring corresponds to the location of the
photon sphere, obtained in the framework of the optical geometry,
which is further associated with the location of the holographic
Einstein ring is fully satisfied with that of the geometrical optics
nicely. Based on our analysis, we argue that holographic images play
a significant role in differentiating the geometric features of
different BHs for the fixed wave source and optical system. Finally,
it is concluded that the holographic images would be interesting to
further characterize the observed image of the BHs in other extended
theories of gravity as well as the implications of this method in
some other fields give us concrete information about the
phenomenological consequences of BHs dynamics. We hope these
observations look bright for the future of the tabletop community.\\

\par \noindent
\par \noindent
{\textbf{Acknowledgments}}\\
\par \noindent
\par \noindent

This work is supported  by the National Natural Science Foundation
of China (Grants  No. $11875095$), Innovation and Development Joint
Foundation of Chongqing Natural Science  Foundation (Grant No.
CSTB$2022$NSCQ-LZX$0021$)  and Basic Research Project of Science and
Technology Committee of Chongqing (Grant No.
CSTB$2023$NSCQ-MSX$0324$).

\end{document}